# Pencil and Paper Electronics: An Accessible Approach to Teaching Basic Physics Concepts

**Pablo Bastante [1] and Andres Castellanos-Gomez[2]**

[1] Departamento de Física de la Materia Condensada, Universidad Autónoma de Madrid, Madrid, E-28049, Spain
[2] Materials Science Factory. Instituto de Ciencia de Materiales de Madrid (ICMM-CSIC), Madrid, E-28049, Spain

E-mail: pablo.bastante@uam.es, andres.castellanos@csic.es

**Abstract**

This teaching article describes a simple and low-cost methodology for studying electrical transport and constructing basic sensor devices using everyday stationery items, including pencils, paper, and a handheld multimeter. The approach is designed for high school and undergraduate teachers and offers an easy-to-implement, hands-on method for teaching fundamental concepts in physical electronics. The materials and experiments outlined in this article are widely accessible and can be easily replicated in various teaching labs, even with limited budgets.

## 1. Introduction

The field of electronics is rapidly expanding, and as a result, there is a growing demand for skilled individuals who can design, build and troubleshoot electronic circuits and components. However, electronics can be an intimidating subject for students who are new to the field, particularly those who do not have access to expensive tools or equipment. As a result, it is important to develop affordable and easy-to-implement practicum experiments that can introduce students to the basics of electronics. By providing hands-on experience with simple electronic circuits, students can gain a better understanding of the underlying concepts and develop the skills they need to succeed in the field.

One interesting approach for creating affordable and easy-to-implement electronic experiments is to use pencil[1–3] and paper[4,5] to create conductive traces on a substrate. This approach is particularly appealing because it does not require expensive tools or materials, and it can be easily adapted to suit a wide range of student skill levels. Pencil-and-paper circuits are also versatile, and they can be used to teach a wide range of topics, including basic circuit theory, sensors, and signal processing.

In recent years, researchers have explored the use of pencil traces on paper to create strain, temperature, humidity, gas and electrophysiology sensors for a wide range of applications.[2,6–14] These sensors have proven to be simple, versatile, and cost-effective, making them ideal for various research projects. These sensors have proven to be simple, versatile, and cost-effective, making them ideal for various research projects. Pencil lead and pencil traces have even attracted the interest for physics education with a variety of interesting experiments dealing with diamagnetism, non-linear heating or the Hall effect.[15–18]

Our work aims to translate this built-up knowledge from research labs to the classroom, demonstrating how these sensors can be used to teach high school and undergraduate students the basic principles of electronics and sensors. By using everyday stationery items such as pencils, paper, and a handheld multimeter (Figure 1), we present an easy-to-implement, hands-on approach for teaching basic physical electronics concepts that is accessible and reproducible in a variety of teaching labs, even under low-budget conditions. We hope that this proposed set of experiments will encourage more students to pursue careers in electronics and related fields.

## 2. Materials

*Paper substrate*: we will use standard office printer paper (usually 80 g/cm2) as the substrate for the fabrication of the devices discussed in this manuscript.

*Pencil*: we will use soft B-type pencils as a source for graphite to 'draw' graphite films on the surface of the paper. B-type pencils have a high content of graphite that ensures good electrical conductivity. We advise having at least two pencils with very different B grades (e.g. 2B and 6B are easily available) in order to study later on the effect of the graphite content in the sheet resistance.[2,19]

*Ruler and scissors*: standard ruler will be used to help in the drawing of the devices and to determine the deflection of the strain sensors. The scissors will be used to cut the devices out of the standard paper piece and to fabricate the paper-based cantilever for the strain sensors.

*Adhesive tape*: To fix the paper substrates during the sample fabrication and to clamp the strain gauge sensors.

*Coffee mug and a kettle*: These items will be needed for the temperature sensor measurement.

*Multimeter*: along this manuscript, we have used a BOLYFA BF117 handheld multimeter that allows connection to an external computer to log the measured data. But any kind of handheld multimeter could be used to reproduce the experiments described here. A multimeter with input for thermocouple temperature sensors would be preferable to calibrate the temperature changes in the practicum with the temperature sensor.

*(Optional) Standard office printer*: A printer can be used to print out the outlines of the devices we plan to fabricate to facilitate drawing the graphite pencil traces.

*(Optional) Magnets and double sided tape*: To apply a stable force to the strain sensors, there have been used magnets with similar masses, fixed to the end of the cantilever by the use of double sided tape.

*(Optional) Wire and Copper and Kapton tapes*: In this work, we have fixed the contact to the devices with wires attached to the electrodes with Copper tape, and the thermocouple has been attached to the mug with Kapton tape.

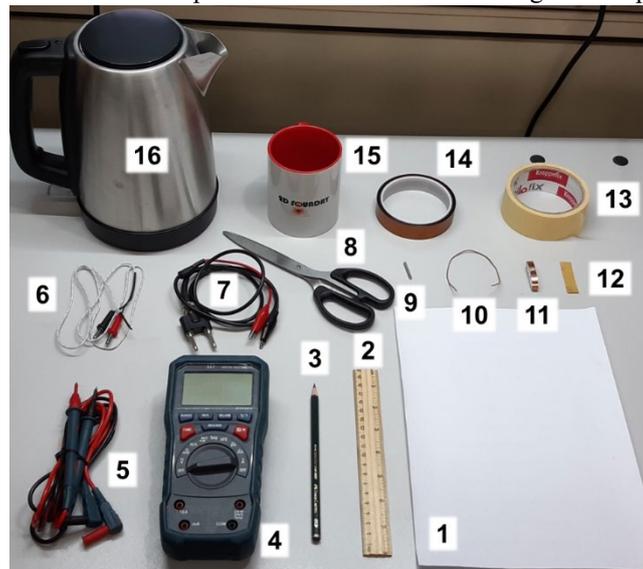

**Figure 1. Materials.** 1 Paper sheet. 2 Ruler. 3 Pencil. 4 Multimeter. 5 Probes. 6 Thermocouple. 7 Crocodile cable. 8 Scissors. 9 Magnets. 10 Wire. 11 Copper tape. 12 Double sided tape. 13 Masking tape. 14 Kapton tape. 15 Mug. 16 Kettle.

## 3. Experiments

*3.1 Sheet resistance measurement*

In this proposed experiment we will determine the sheet resistance of films of graphite on paper.

*3.1.1 Preparations*

We use a regular office printer to print some guiding lines on a piece of paper (see Figure 2a). We will use these guiding lines to draw our graphite film with the help of a masking tape and the selected B-type pencil on paper (see Figure 2b-e), and to find the positions to place the multimeter probes (see Figure 2f). Note that these guiding lines can be also hand-drawn with the help of a ruler if a printer is not available.

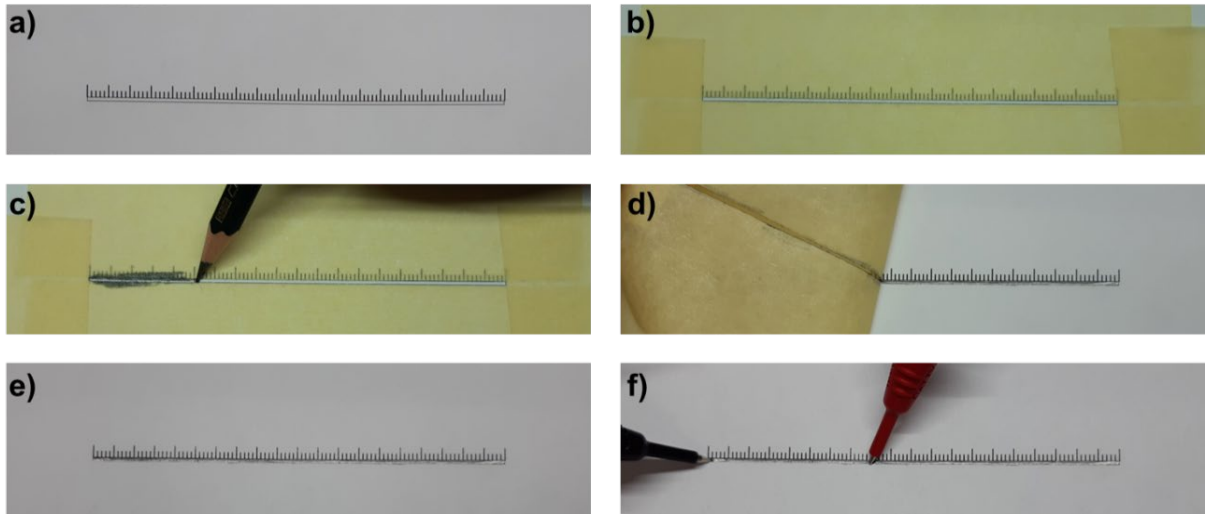

**Figure 2. Sheet resistance sample.** (a) The outline of the conductive trace, a mm scale bar for length characterization, is printed on a paper sheet. (b) The masking tape is placed on the outline as a mask. (c) The outlines are drawn using pencils with different hardness. (d) The mask is removed. (e) A picture of the final sample. (f) Demonstration of the measurement contacting with the probes of the multimeter at one edge of the sample and one of the marked points every 5 mm.

*3.1.2 Experiment*

The students will measure the resistance by placing the multimeter probes at points separated by longer and longer distances.

*3.1.3 What do we expect to observe*

For an ohmic material one would expect to see a linear dependence of the resistance with the distance between the probing tips. Given the fact that the graphite trace is formed by a network of interconnected graphite flakes, one could observe deviations with respect to the ideal linear trend. But in some cases, where a dense graphite trace is drawn the dependence follows accurately the expected ohmic dependence. By fitting the data to linear regression (Figure 3), one can obtain the contact resistance (resistance to the flow of electrical current due to the non-perfect contact between the surfaces of the channel material, graphite, and the electrodes, the metal probes of the multimeter) [20] as the intercept, and the slope is proportional to the square resistance [21]:

$$esistance = \frac{\rho}{t}\frac{Distance}{Width} = \frac{R_\square}{Width} Distance$$

where ρ is the material resistivity and t is the sheet thickness, so the square resistance is defined as $R_\square = \rho / t$. In this work, the width of the channel used is 1 mm.

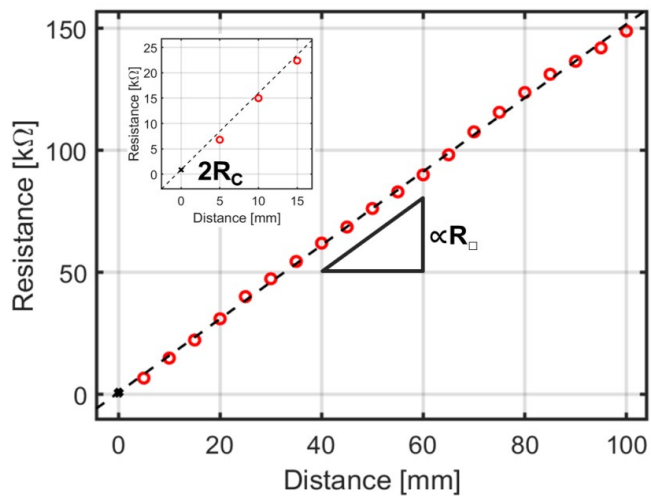

**Figure 3. Transfer length method.** Distance dependence of the resistance measured in one sample made with 4B pencil. The measured points in red are disposed of in a linear distribution. The data is fitted to linear regression, shown as a black dashed line. The slope of the regression is proportional to the square resistance ($R_\square$), and the intercept is twice the contact resistance ($R_C$). (inset) Zoom close to the origin of the coordinates to elucidate the offset in resistance at zero distance, representing the contact resistance.

We advise the students to measure the resistance vs length on 10 different graphite traces to get information about the sample-to-sample statistical deviations. Repeating the experiment with different B grade pencils would be very interesting (Figure 4a). Lower B grade pencils have a lower content of graphite. Thus comparing harder pencils (e.g. HB and 2B) with soft ones (e.g. 6B or 8B) lead to a noticeable difference in the sheet resistance (Figure 4b). This sheet resistance dependence on the hardness of the pencil is in good agreement with previously reported works that correlate it with the content of graphite in the pencil lead: higher graphite content for softer pencil leads.[2,19] However, it is imperative to acknowledge the stochastic characteristics inherent to the network architecture formed by interconnected graphite flakes deposited by the same pencil (as illustrated in the inset of Figure 4b). Within such systems, electrical conduction primarily occurs via a hopping mechanism, wherein charge carriers traverse between overlapping flakes until a percolative pathway is established. Consequently, even minor fluctuations in the density of these flakes can exert a pronounced impact on the availability of percolative pathways for current propagation. This subtle variation can be observed in measurements taken from different samples, despite using the same pencil composition consistently."

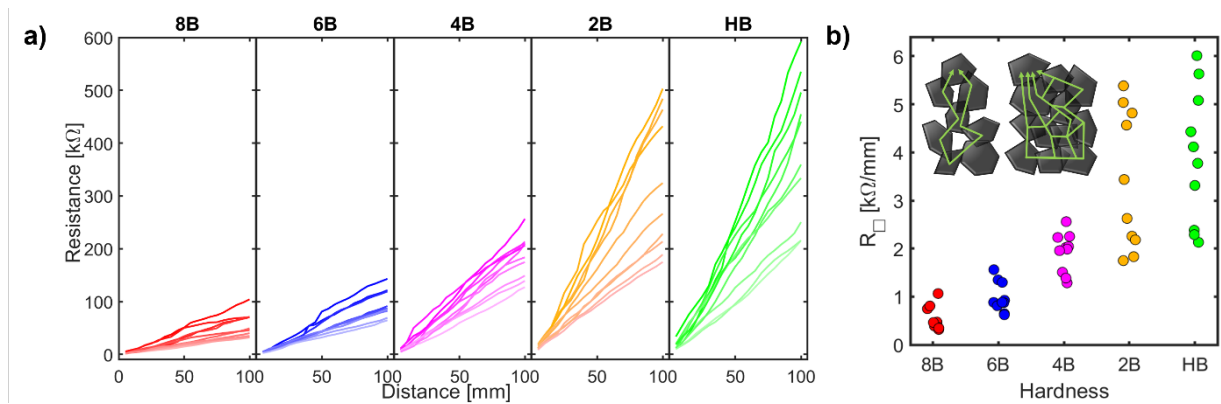

**Figure 4. Pencil hardness dependent sheet resistance.** (a) Distance dependence of the resistance for different hardness pencils from HB to 8B, measured in 10 samples for each hardness. (b) Hardness dependence of the square resistance ($R_\square$)

calculated as the slope of the linear regression of the distance dependence of the resistance for each sample. (Inset) Representation of the overlap and the conductive paths in two networks with different density of interconnected graphite flakes.

*3.2 Strain sensor*

In this proposed experiment we will fabricate a very simple strain gauge: a device that transduces an external strain to an electrical signal.

*3.2.1 Preparations*

Similar to previous experiment, we will start with a piece of standard office printer paper where we will print out the outline of the device (Figure 5a). The device will consist of a paper based cantilever with a graphite trace meander near the clamping point of the cantilever. After drawing the graphite trace, the device is connected with wires attached to the electrodes by the use of copper tape (Figure 5d). Notice that the electrodes can be contacted directly with the probes of the multimeter.

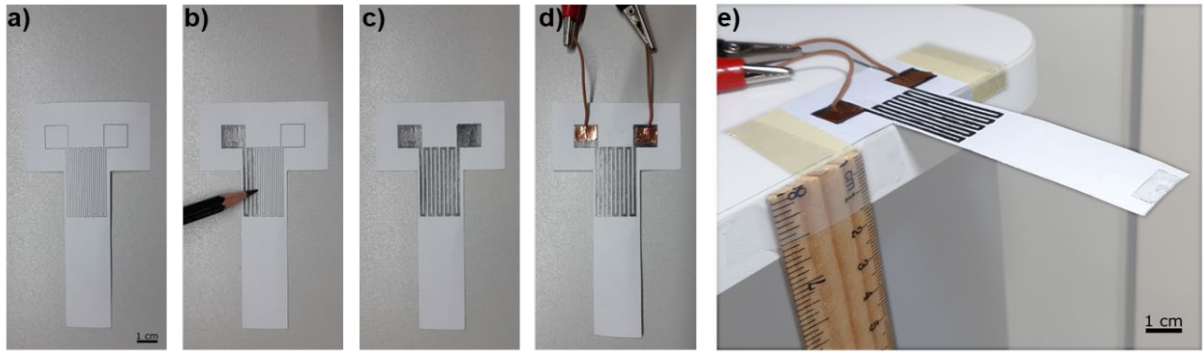

**Figure 5. Strain sensor fabrication.** (a) The outline of the device is printed on a paper sheet. (b) The outlines are drawn using a 4B hardness pencil. (c) A picture of the final sample. (d) The sample is connected with wires contacting by the use of copper tape and connected to the multimeter with the use of crocodile cables. (e) The device is attached to the edge of the table, with double sided tape at the end to stick the masses. The ruler is attached near the device to measure the deflection.

*3.2.2 Experiment*

For the measurements the base of the cantilever will be firmly clamped to the edge of a desk using adhesive tape. The ruler is clamped nearby to quantitatively measure the bending of the cantilever to correlate the resistance of the graphite trace with the displacement of the cantilever (Figure 5e). The resistance of the graphite trace will be monitored with a handheld multimeter while another student bends the cantilever upwards and downwards.

*3.2.3 What do we expect to observe*

Upon upwards bending one would expect a decrease of the resistance while bending downwards would lead to an opposite trend. This is due to the nature of the graphite trace films, composed of a network of interconnected flakes. A compressive strain yields an increased flake-to-flake overlap, thus reducing the resistance. Tensile strain, on the other hand, leads to reduced overlap and increase resistance.[6,22,23] Another proposed experiment is to load test masses at the end of the cantilever (e.g. paper clips or small magnets) as depicted in Figure 6a and monitor the change of resistance upon mass loading (Figure 6b). The Strain is calculated from the recorded deflection ($d_F$) as [24]:

$$Strain = \frac{3 d_F h (x_F - x_s)}{2 x_F^3}$$

where h is the thickness of the substrate (the paper sheet), $x_F$ is the distance from the clamping point to where the force is applied (or the mas is loaded) and $x_s$ is the distance from the clamping point to the middle of the graphite trace along where the strain is applied. The deflection is then determined as the vertical distance between the clamping point and the end of the device, ie, where the mass is loaded. This value is experimentally measured by acquiring front-view photographs of the cantilever with a ruler which allows us to directly convert the dimensions directly measured in pixels in the digital picture to distance units. From fitting the differential resistance (defined as the change in resistance normalized by the initial resistance value) vs. Strain data to linear regression, it is obtained the Gauge factor (defined as the ratio between the change in resistance and the mechanical strain) as the slope (Figure 6c). This is a very simple demonstration of a mass sensor using a strain gauge to transduce the change in mass load to an electrical signal.

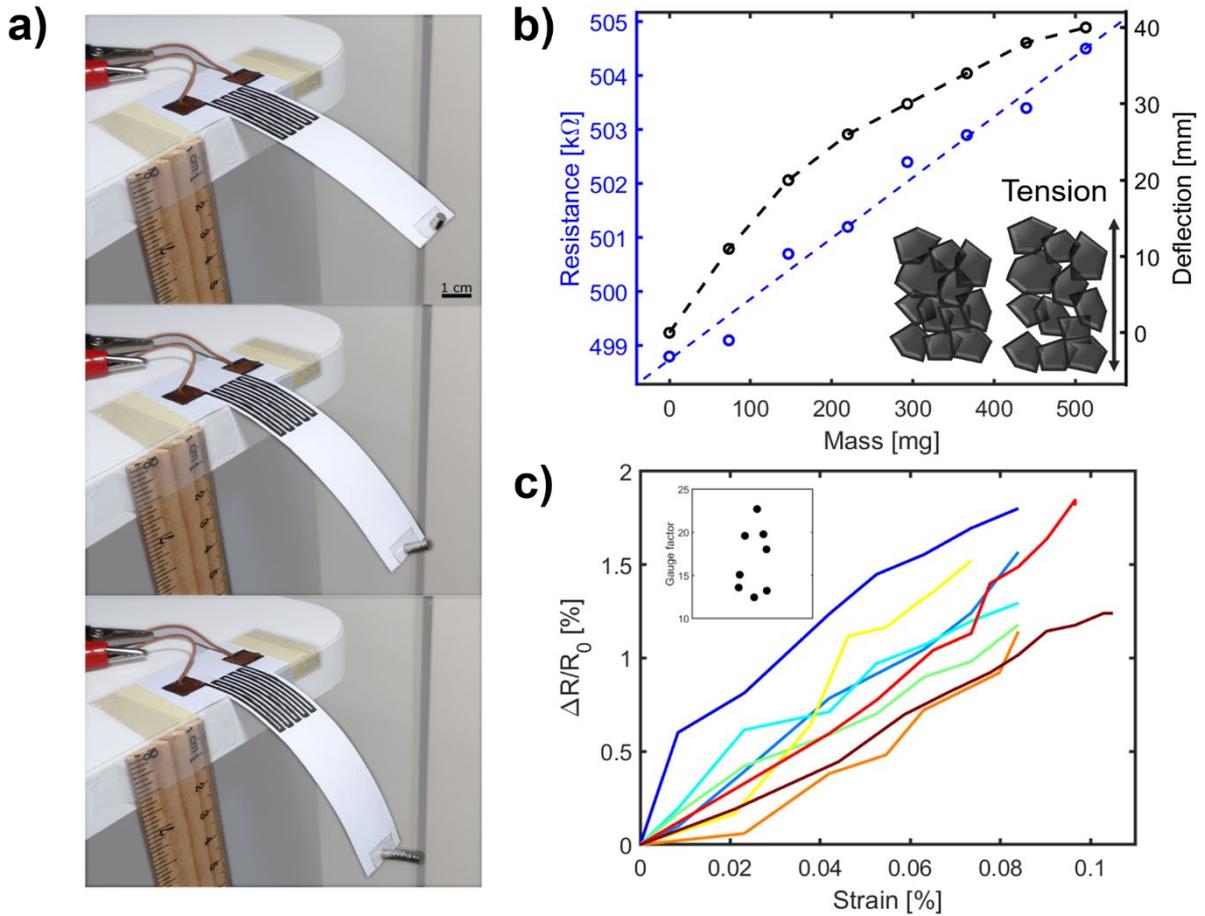

**Figure 6. Strain sensing.** (a) Images of the device loaded with an increasing number of masses, thus increasing the deflection. (b) Mass dependence of the resistance, in blue with the dots as the data points and the linear regression as a dashed line, and the deflection, which describes a nonlinear dependence. (Inset) Representation of the overlap in the graphite network and the changes with uniaxial strain. (c) Strain dependence of the differential resistance for 8 devices to illustrate the device-to-device variability which is expected for this kind of systems composed of a networks of interconnected flakes randomly distributed during the drawing process. (inset) Gauge factor of each sample calculated as the slope of the linear regression of the strain dependence of the differential resistance data.

*3.3 Temperature sensor*

In this proposed experiment we will implement a simple temperature sensor based on a thermistor, i.e. a device whose electrical resistance depends on the temperature.

*3.3.1 Preparations*
Using a geometry similar to the one used for the strain sensor, we draw a meander-like graphite trace on paper.

*3.3.2 Experiment*
To test the temperature dependence resistance of the graphite trace we propose to attach the graphite trace on paper device to a coffee mug and register the resistance of the film as a function of time. Then pouring hot water from a kettle (Figure 8a) and monitor the changes in the resistance. A thermocouple is attached to the mug with Kapton tape close to the device to monitor temperature and correlate it with the resistance changes (see Figure 7c).

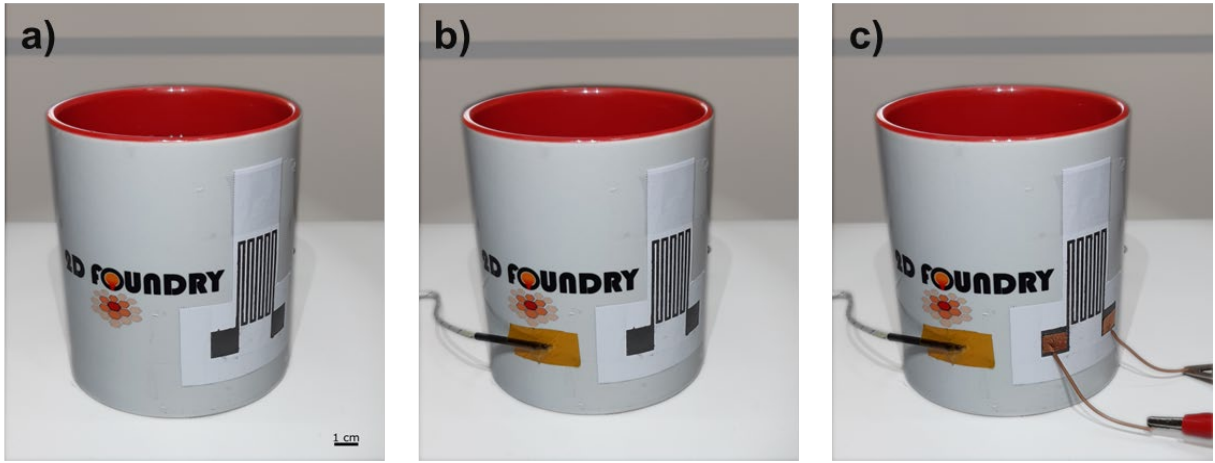

**Figure 7. Temperature sensor fabrication.** (a) The sensor is attached to the mug. (b) A thermocouple is attached to the mug near the sensor with Kapton tape and connected to the multimeter. (c) The sensor is connected with wires contacting by the use of copper tape and connected to the multimeter with the use of crocodile cables.

*3.3.3 What do we expect to observe*

Given the fact that graphite is a semi-metal, it is very easy to thermally excite charge carriers from the valence band to the conduction band, leading to a decrease of the resistance. The resistance vs. Time trace resembles the temperature vs. time one measured with the thermocouple (Figure 8b). The temperature dependence of the resistance describes a linear behavior (Figure 8c). From fitting the data point to linear regression, it is obtained the temperature coefficient of resistance. We obtain temperature coefficient of resistance values that span from -2500 ppm/°C to -3700 ppm/°C, in good agreement with other graphite on paper temperature sensors reported in the recent research literature.[8,14,25]

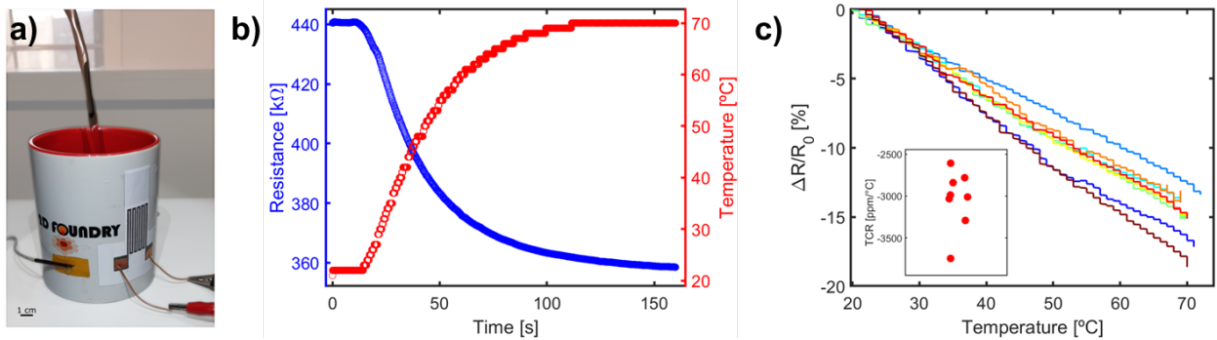

**Figure 8. Temperature sensing.** (a) Demonstration of the measurement procedure, dropping hot water in the mug with the sensor and the thermocouple attached to it. (b) Time dependence of the resistance and the temperature after dropping the hot water in the mug. (c) Temperature dependence of the differential resistance $\Delta R/R_0$ of the sensor for 8 different sensors. (inset) Temperature coefficient of resistance (TCR) for each sensor calculated as the slope of the linear regression of the curves in the figure.

## Conclusions

The proposed methodology utilizes everyday stationery items to teach basic physical electronics concepts and fabricate simple sensor devices. By using easily available and inexpensive materials, this approach makes it possible to reproduce many of the proposed experiments in high school and undergraduate teaching labs, even under low budget conditions. This methodology emphasizes hands-on experimentation and data analysis, providing students with a deeper understanding of electrical transport and sensor technology. The experiments described in the article can be adapted to suit different academic levels, making it a versatile and valuable teaching resource.

**Supporting Information**

Supplementary Information includes videos demonstrating the working principle of the experiments in real time.

**Funding Sources**

European Research Council (ERC) through the project 2D-TOPSENSE (GA 755655)
"FLAG-ERA program (JTC 2019) under the project To2Dox (PCI2019-111893-2)
European Union's Horizon 2020 research and innovation program (Graphene Core2-Graphene-based disruptive technologies and Grant agreement No. 881603 Graphene Core3-Graphene-based disruptive technologies)
Comunidad de Madrid through the project CAIRO-CM project (Y2020/NMT-6661)
Spanish Ministry of Science and Innovation through the projects PID2020-115566RB-I00.

**Acknowledgements**

This work was funded by the European Research Council (ERC) under the European Union's Horizon 2020 research and innovation program (grant agreement n° 755655, ERC-StG 2017 project 2D-TOPSENSE), the European Union's Horizon 2020 research and innovation program (Graphene Core2-Graphene-based disruptive technologies and Grant agreement No. 881603 Graphene Core3-Graphene-based disruptive technologies), the Ministry of Science and Innovation (Spain) through the projects PID2020-115566RB-I00. We also acknowledge funding from the "FLAG-ERA program (JTC 2019) under the project To2Dox (PCI2019-111893-2) and the Comunidad de Madrid through the CAIRO-CM project (Y2020/NMT-6661)

**Data Availability Statement**

The data that support the findings of this study are available from the corresponding author upon reasonable request.